# Temperature dependence of a graphene growth on a stepped iridium surface

Iva Šrut[*], Vesna Mikšić Trontl, Petar Pervan, Marko Kralj

Institut za fiziku, Bijenička 46, HR-10000 Zagreb, Croatia

**Abstract**

We have used scanning tunneling microscopy to study the growth of graphene on a periodically stepped Ir(332) substrate surface, which is a promising route for modification of graphene properties. We have found that graphene continuously extends over iridium terraces and steps. Moreover, new distinctive mesoscopic features of the underlying surface are formed involving large, flat terraces accompanied by groups of narrower steps. The distribution of the newly formed terraces is sensitive to the preparation temperature and only below 800°C terrace width distribution closer to the intrinsic distribution of clean Ir(332) are found. We propose that the microscopic shape of steps found after graphene formation is strongly influenced by the orientation of graphene domains, where graphene rotated by 30° with respect to the substrate has a prominent role in surface structuring.

---
[*] Tel/Fax: +385 1469 8889, E-mail address: isrut@ifs.hr,

## 1. Introduction

Graphene is a promising material for future application in nano-electronic devices [1, 2]. The extensive research of epitaxial graphene on different substrates is an open route to the large-scale applications of graphene. Graphene can be grown on transition metals by a chemical vapor deposition (CVD) technique [3], and it can be transferred from a metal to other supports, e.g. to polymer sheets to make a flexible transparent touch screen display, where graphene has a role of a transparent electrode [4]. Possibility of such manipulation makes CVD grown graphene attractive for a number of applications.

Many of the desirable properties of graphene are related to its electronic structure with conical $\pi, \pi^*$ bands (Dirac cone) of vertices touching in a single point at the Fermi level. This makes graphene a gapless semiconductor which is a limiting factor for some applications of graphene in electronics where a sizeable band gap at the Fermi energy is required. For that reason, a lot of effort is directed towards manipulation of the electronic structure of graphene around the Fermi energy. One possible route, which is subject of this work, relies on altering the graphene band structure by means of an additional periodic potential, i.e. graphene superlattice. Calculations indicate that the lateral superlattice structures may lead to unexpected and potentially useful charge carrier behavior, e.g. gap openings or Fermi velocity anisotropy [5]. Graphene superlattices have been realized on well defined single crystal substrate surfaces as a consequence of the mismatch of graphene and substrate lattices which leads to the so called moiré superstructures. For graphene on Ir(111) it has been shown that the moiré superstructure adds a long range superperiodic potential to graphene, which opens minigaps in the Dirac cone [6]. For the same system it was demonstrated that the adsorption of hexagonal array of metal-clusters gives rise to highly anisotropic Dirac cones [7].

A one dimensional (1D) type of graphene superlattice can be achieved by adding a 1D periodic potential to graphene, in which context it is appealing to consider a periodically stepped metal surface as a substrate for the growth of graphene. Motivated by this consideration, in our work we used scanning tunneling microscopy (STM) to explore structures of graphene prepared under various growth conditions on a stepped Ir(332) surface. Iridium was chosen as a substrate because Ir(111) is known to support growth of single layer graphene which can be tuned to uniform orientation over the entire sample surface [8, 9], extending across step edges [10]. For many metals the issues of uniform thickness and varying graphene rotation with respect to the substrate are often limiting factors for graphene quality at large scales. The uniform orientation cannot be achieved e.g. on foils due to their polycrystallinity [11] or on a single crystal Pt(111) due to very weak binding of graphene [12]. On Ir(111), macroscopically aligned graphene is achieved only by the careful optimization of the growth procedures [6, 8], whereby the substrate temperature turns to be the main growth parameter. The temperatures above 1230°C are needed to grow graphene aligned to the substrate with perfectly matched graphene $\langle 11\bar{2}0 \rangle$ and Ir $\langle 10\bar{1} \rangle$ directions, which is referred to as the R0 [13], implying 0° rotation. For this phase, approximately 10% difference in lattice parameters of graphene and Ir results in an incommensurate hexagonal (9.32×9.32) moiré superstructure with a repeat distance of 2.53 nm [6]. Thus the R0 moiré unit cell consists of about 200 carbon atoms positioned differently with respect to atoms of an Ir lattice. With respect to the carbon atom stacking, the moiré cell is prominently divided into atop, fcc, and hcp regions [14]. The atop-type region, where the iridium atom lies right in the centre of the carbon ring, defines the corners of the moiré cell [14]. For lower growth temperatures, graphene on Ir(111) grows in many minor different orientations where specific orientations of 14°, 18.5° and 30° appear along side with 0° [13]. Most frequent minor contribution is of 30° oriented phase, i.e. with 30° rotated graphene $\langle 11\bar{2}0 \rangle$ and Ir $\langle 10\bar{1} \rangle$ directions, which is referred to as the R30, and has a moiré repeat distance of 0.5 nm [13, 9]. It should be noted that besides lattice mismatch, which is fixed by the choice of the substrate, the relative rotation of graphene with respect to the substrate leads to different sizes of moiré unit cells, with the R0 having the largest repeat distance.

Our study indicates that the growth of graphene on a stepped Ir(332) surface alters its original narrow terrace width distribution (TWD). In terms of surface morphology, our result show the formation of two different areas: (i) compact areas with narrower step spacing (one to three atoms wide) wide up to 15 nm, so called step bunches, and (ii) areas consisting of single flat (111) terraces up to 19 nm wide. We find that graphene extends across step edges in the shape of domains of different orientations and sizes, ranging up to 100 nm. The detailed analysis indicates that graphene domains consisting of the R30 graphene result in a morphology with pronouncedly straight Ir step edges.

## 2. Experimental

The experiments were performed in two ultra high vacuum (UHV) setups. Both UHV systems were operated at a base pressure of $5\times10^{-10}$ mbar. One setup was used for low energy electron diffraction (LEED) measurements, and the other for the STM characterization of clean and graphene covered Ir(332) surface. The STM measurements were performed at room temperature. The STM was calibrated by measurements on the HOPG sample and the STM images were processed by a WSxM software [15]. The applied bias voltage to the sample was in the range of 130 mV to 300 mV, while the tunneling current was around 0.5 nA for all images except for the atomically resolved STM image in Fig. 2a ($I_t$=2.01 nA). Same Ir(332) sample was used in both setups, a crystal 6 mm in diameter, of 99.99% purity, polished with roughness <0.03 μm and with an orientation accuracy <0.1°. In both setups the sample was heated by e-beam heating using a hot filament at negative potential near to the grounded sample. In the system for LEED measurements a C type thermocouple was directly connected to the sample, whereas, in the STM system, sample temperature was calibrated by the C type thermocouple attached to the sample with respect to the heating time and power used and, additionally, a values measured by a K type thermocouple attached at the sample plate. The C type thermocouple was then removed from the sample to enable multiple graphene preparations and transfer from the sample preparation manipulator to STM.

2.1 Substrate cleaning

Ir(332) sample was cleaned by using cycles consisting of 50 min of 0.8 keV argon-ion sputtering while the sample was kept at 200°C, followed by 10-25 min of annealing in oxygen partial pressure of $1\times10^{-7}$ mbar at 800°C, and a final 5-15 min annealing in UHV at 850°C. Each cycle was finished by uniformly cooling the sample to room temperature over a period of 30-40 min. After five such cycles typically a sharp

LEED pattern and large scale STM images confirmed clean surface with ordered array of steps.

2.2 Graphene growth

Graphene preparation methods used in this work involved two well known procedures. One is a room-temperature hydrocarbon adsorption followed by decomposition at a fixed, elevated temperature, referred as a temperature-programmed growth (TPG). Another procedure, CVD, consists of exposing the hot substrate to hydrocarbon gas. Growth conditions (temperatures, pressures, hydrocarbon doses) described below were used by taking known conditions for preparation of graphene on Ir(111), but not crossing 1120°C to avoid irreversible structural transformation of the nominal stepped surface. Note that this condition limits us to temperatures below 1230°C needed to achieve single domain of the R0 graphene [13, 14].

We have prepared graphene either by (i) CVD or by (ii) one TPG cycle followed by CVD. In this work, we have focused to the correlation of a temperature used during CVD cycle and a resulting graphene structures. We did not notice any significant influence of the hydrocarbon pressure parameter in the range $5\times10^{-9}$ to $5\times10^{-8}$ mbar, used in our work. Preparation (i) was performed at two sample temperatures, 775°C and 880°C, with corresponding ethylene pressures and duration presented in Table 1. Preparation (ii) involved dosing the ethylene at the pressure of 2×10$^{-8}$ mbar for 30 s, followed by annealing to high temperatures which corresponded to the ones of subsequent CVD cycles. The CVD was performed at three different sample temperatures, 770°C, 820°C and 930°C (see Table 1). Graphene growth was finalized by short annealing at the CVD temperature and slow cool down (30-45 min) to room temperature. The combined TPG+CVD preparation was motivated by findings that such procedure produces a uniform orientational order of graphene on flat Ir(111) surface [8].

| preparation method | temperature [°C] | pressure [mbar] | time [min] | corresponding figure |
|---|---|---|---|---|
| **CVD** | 880 | 5x10$^{-8}$ | 21 | 1a |
|  | 775 | 1.5x10$^{-8}$ | 12 | 1c |
| **TPG+CVD** | 770 | 3x10$^{-8}$ | 25 | 3a |
|  | 820 | 2x10$^{-8}$ | 20 | 3b |
|  | 930 | 5x10$^{-9}$ | 90 | 3c |

**Table 1.** Overview of preparation parameters used in this work with a reference to figures with STM images.

## 3. Results

3.1 Clean Ir(332)

Structural model of an ideal Ir(332) surface is displayed in Fig. 1a. The model shows (111) terraces 1.25 nm wide, consisting in width of $5\frac{1}{3}$ atoms, and (111) steps with edges parallel to the $\langle 10\bar{1} \rangle$ high-symmetry direction. The step height of 0.22 nm corresponds to the interlayer distance between the

(111) planes of iridium. The LEED pattern of clean Ir(332) surface, Fig. 1b, reproduces a hexagonal atomic arrangement of the surface. Also the additional splitting of diffraction spots is visible. This, so called, spot splitting reflects the periodicity of ordered array of steps corresponding to (1.3±0.1) nm, matching the expected value.

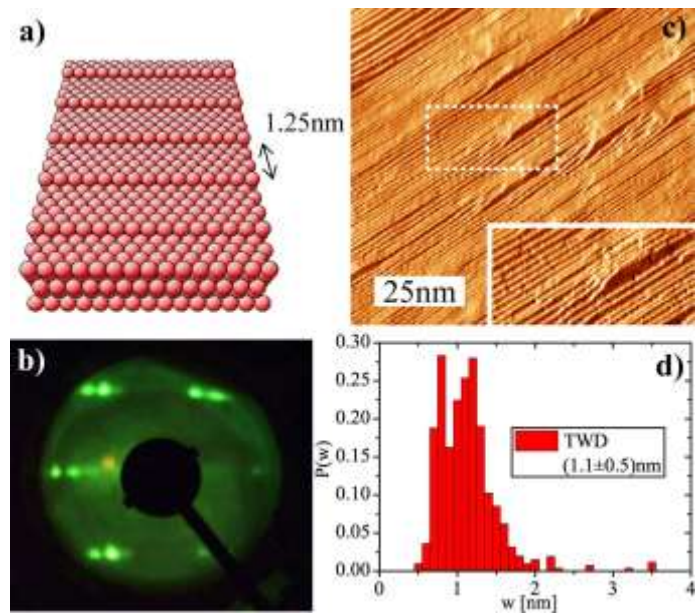

**Figure 1.** a) Structural model of a (332) fcc-crystal surface generated by Surface Explorer software[1]. b) LEED pattern taken at the electron energy of 64 eV, c) STM image of clean Ir(332) with an inset showing 1.5 times magnified region marked by a white dashed rectangle, and d) TWD of clean Ir(332).

Fig. 1c, which shows a large scale STM image, confirms that the Ir(332) surface was well ordered and clean. Inset of Fig. 1c illustrates that small local deviation in terrace widths may be

---
[1] http://surfexp.fhi-berlin.mpg.de

seen after cleaning process. Analysis of several large scale STM images taken at different surface positions gives a narrow TWD with a mean value of (1.1±0.5) nm (Fig. 1d). TWD is shown as the probability P(w) of finding terrace of the width w. Note a small discrepancy between the mean width values obtained by STM and LEED. This can be attributed to the local variation of the terrace width e.g. due to the finite processing accuracy in crystal orientation of <0.1°. Since STM is a local technique it will be more susceptible to local variations than LEED, which averages over macroscopic sample scales.

3.2 Graphene growth and morphology

Graphene covered areas were easily identified by observing the honeycomb carbon lattice (with unit cell parameter of 2.45 Å) and, at larger scales, the moiré structure. Both structural features are seen in Fig. 2a, where a moiré unit cell is marked by a white rhombus. Carbon rows of graphene are rotated by 41.2° with respect to the moiré lattice, as marked in Fig. 2a. This rotation serves as a „magnifying lense" to determine graphene rotation with respect to iridium, and by using formalism developed in Ref. [14] we calculate it in this case be 5.7°±0.1°. This is in a good agreement with the measured moiré unit cell of (1.80±0.05) nm which corresponds to the rotation of 5.8°±0.4° [14].

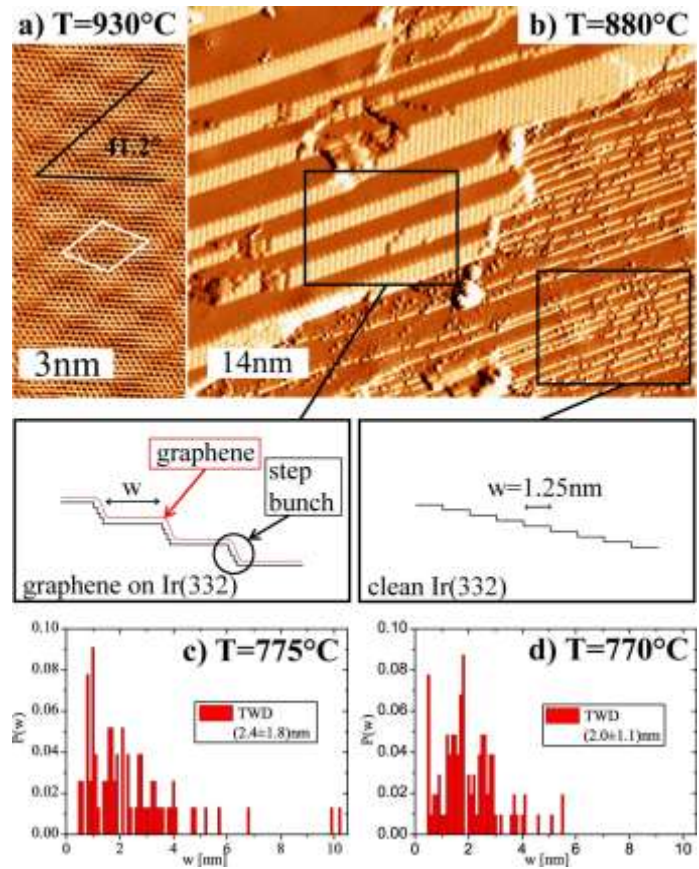

**Figure 2.** a) STM image of graphene area showing atomic and moiré resolution. b) STM image of a Ir(332) surface partially covered by graphene. Panels below are schematic models of graphene covered and bare Ir(332). TWD of c) graphene grown by CVD method at 775°C and d) graphene grown by TPG+CVD method at 770°C.

Fig. 2b shows an STM image of Ir(332) partially covered with graphene. Graphene covered areas obviously have different morphology compared to clean Ir(332), which is readily seen in Fig. 2b and schematically illustrated in the panels below it. Graphene growth leads to the formation of a number of wide (111) terraces followed by groups of narrower steps (one to three atoms wide), i.e. a step bunch. Graphene grows continuously over terraces and step bunches. This can be confirmed by analyzing the orientation and the moiré repeat distance of graphene which extends over several terraces and step bunches. In Fig. 3c inset one such analysis was done for moiré with a repeat distance of 1.36 nm. In this work we have observed that TPG+CVD growth can lead to a narrower TWD compared to growing graphene just by CVD, with the mean terrace width closer to the value of the clean Ir(332) surface (cf. Fig. 2c, 2d). Note that our TWD data in Figs. 2 and 3 include only the widths of the wide (111) terraces, and not of the very narrow steps in step bunches.

3.3 Temperature dependence of graphene growth

Since the motivation for our work is the modulation of the electronic structure of graphene by uniaxial periodic structure-induced potential, in the remaining part of the paper we focus to TPG+CVD growth procedures with narrower TWD's. In particular, we focus to several different temperatures during the CVD cycle: 770°C, 820°C and 930°C, presented by large scale STM images in Fig. 3a, 3b and 3c, respectively. One can observe a general feature that the (111) terraces are getting wider as the preparation temperature increases.

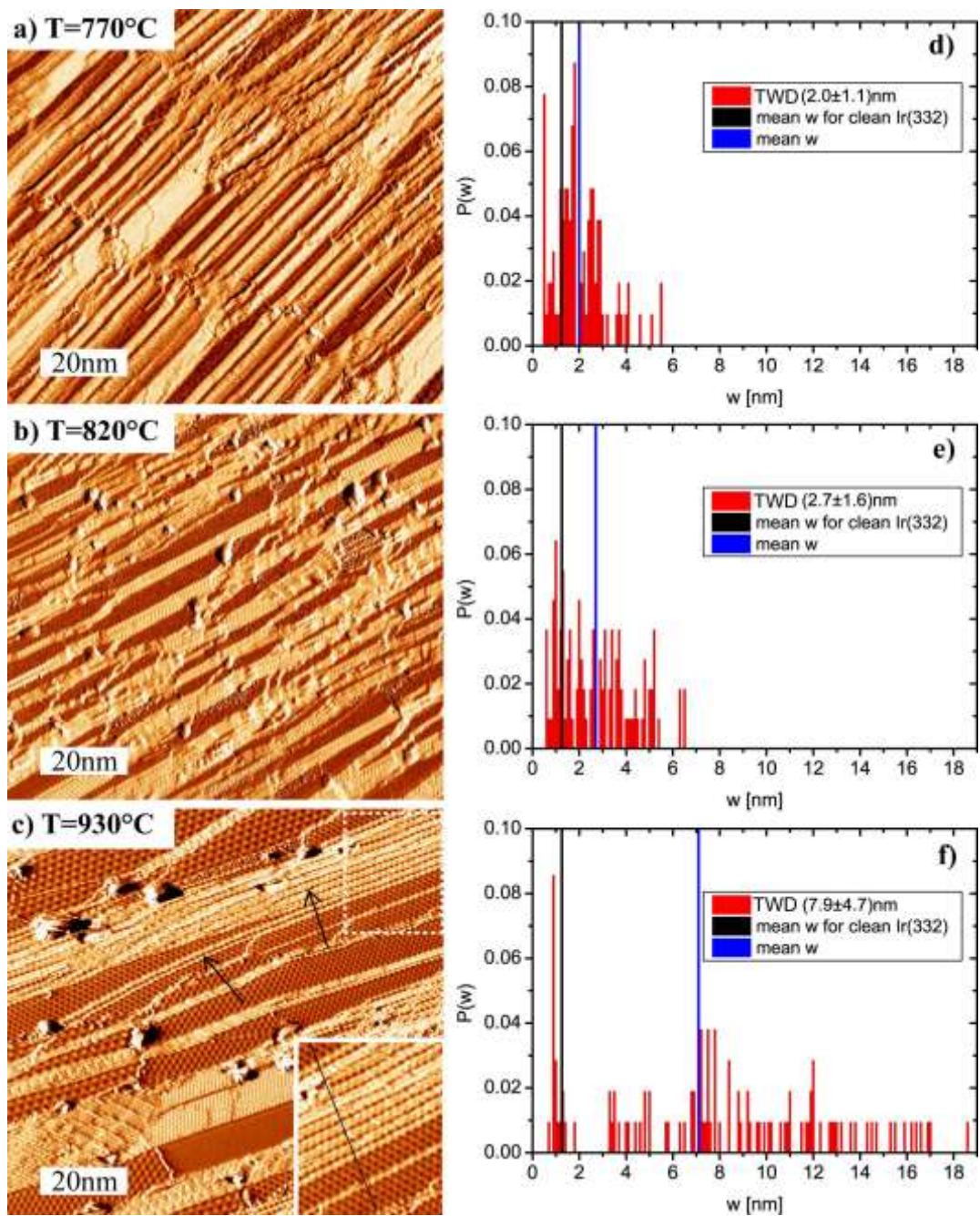

**Figure 3.** STM images of graphene covered Ir(332) for different preparation temperatures (a-c), and corresponding TWD's (d-f). Inset in c) shows 1.5 times magnified region marked by white dashed rectangle. The black line corresponds to the diagonal direction across the moiré unit cell.

In Figs. 3d, 3e, 3f, this is clearly seen from TWD's for corresponding temperatures. The mean terrace width shifts from 2.03 nm for 770°C to 7.91 nm for 930°C. In addition to this, TWD data also reveal significant smearing of distributions for higher preparation temperature. The smearing effect enables us only to compare TWD's in terms of overall range and mean terrace widths. Pronounced peaks slightly below and around 1 nm are seen in TWD's for all preparation temperatures and are due to a large number of narrow terraces, examples of which are marked by arrows on Fig. 3c.

3.3 Analysis of rotational domains and moiré structures

The LEED pattern in Fig. 4a corresponds to the sample prepared under the same conditions as the one shown in STM image in Fig. 3c. The diffraction pattern reveals that the surface is covered by graphene in several different orientations. Dominant carbon diffraction intensities are grouped around 0° orientation (R0) with an angular spread of ±6°, and around 30° orientation (R30) in a similar angular spread but with distinctive spots at symmetric 26° and 34° (R26) positions. In addition, a faint diffraction ring is visible, suggesting also any other possible orientation of graphene. The diffraction maxima corresponding to Ir(111) are now visible in a hexagonal pattern but without characteristic spot splitting of uniformly ordered stepped surface. This is consistent with rather wide TWD from Fig. 3f.

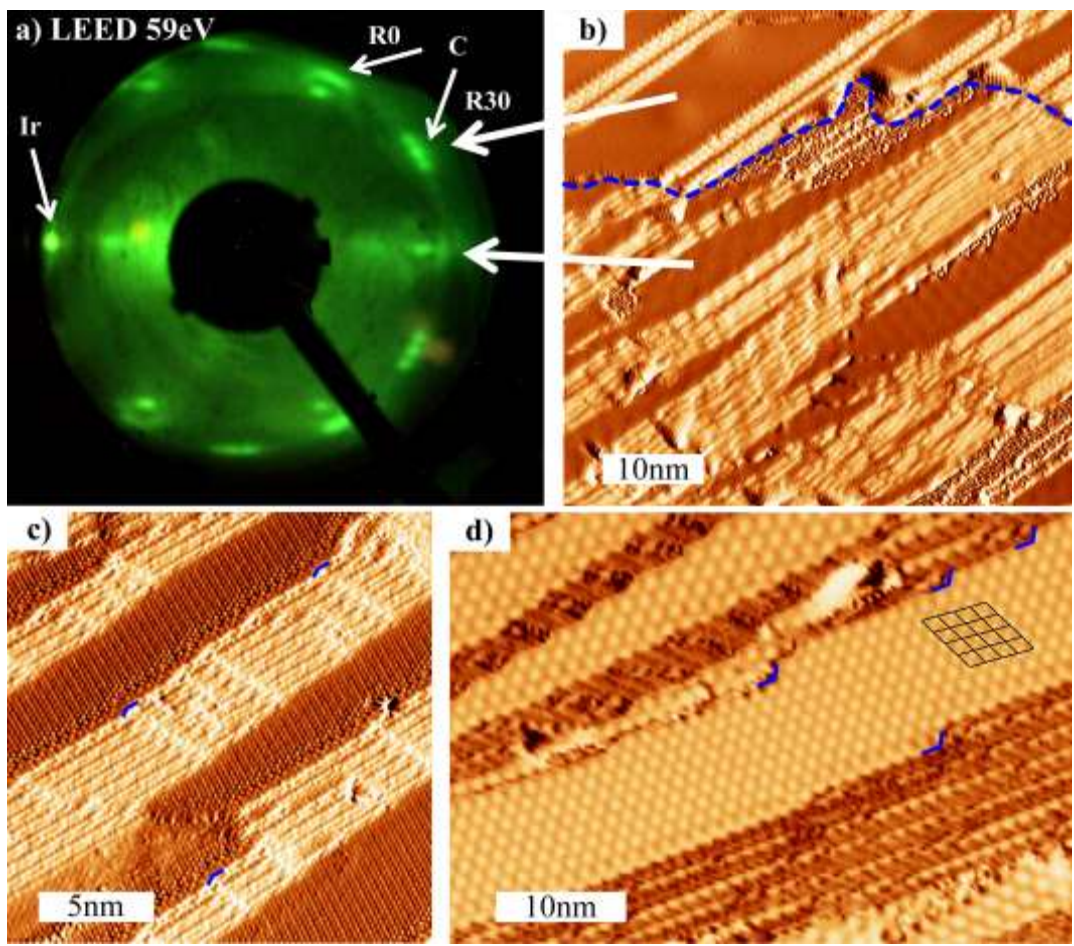

**Figure 4.** a) LEED pattern taken at the electron energy of 59 eV, after growing graphene by TPG+CVD method at temperature of 930°C. b) STM image of two connected rotational domains of graphene corresponding to R0 and R30 groups. c) STM image of the R30 graphene showing atomic resolution at terraces and step bunches where three kinks are marked by blue lines. d) STM image showing several domains within the R0 group graphene with clearly visible moiré patterns and four kinks marked by blue lines.

Similar to 930°C, the LEED pattern for preparation temperature of 1120°C (not shown) reveals moiré pattern around the R0

graphene spot and a faint but visible contribution at the R30 and the R26.

LEED patterns provide us with relevant information on the surface structure averaged over macroscopic areas. In addition, STM enables us to characterize microscopic areas of certain orientation. The analysis of the moiré pattern is especially helpful in that sense (cf. Fig. 2a). Moreover, STM images can be used to characterize the appearance of underlying Ir steps and step edges. Figure 4b shows the STM image of graphene covered surface with two distinct graphene domains (separated by blue dashed line). The upper section of the image shows graphene domain orientated within the R30 group, while the lower section exhibit graphene domain orientated within the R0 group. It is obvious that graphene in the upper part of the image shows pronouncedly strait step edges with only few kinks, while the lower graphene domain shows distinctly different, less regular step edges with frequent step branching and kinks. The observation that graphene domains around the R30 form straighter step edges, is additionally corroborated by the analysis of other large scale STM images. Figure 4c shows a single R30 domain, and Fig. 4d shows several domains with orientations close to the R0. On both images, the kinks (marked by blue lines) are found at step edges. The kinks on a single graphene domain (cf Fig. 4c) are all of the same size, in a direction perpendicular to the steps.

Additionally, the moiré superstructure is clearly visible in Fig. 4d, where the bright protrusions correspond to the atop regions that corner moiré cells (marked by black net of rhombuses). From Fig. 4d it is evident that the step edges are terminated by these atop regions, and further, that kinks as well occur at the atop positions.

## 4. Discussion

In this work it is demonstrated that it is possible to cover the entire Ir(332) surface by a monolayer graphene. LEED patterns showed spots that reveal the presence of graphene with several different orientations, grouped around the R0 and the R30 rotations (cf. Fig. 4a). LEED patterns also revealed a loss of uniform order of pristine Ir(332) surface through deterioration of characteristic diffraction spot splitting. STM images with atomic and moiré resolution accordingly corroborated the presence of graphene in various orientations that spread over several terraces (cf. Fig. 3c). Moreover, STM characterization clearly revealed mesoscopic restructuring of the surface upon graphene formation accompanied by two new structural features: (i) wide (111) terraces, and (ii) groups of narrower steps - step bunches, consisting of steps one to three atoms wide. Such step bunches do not occur on perfectly clean Ir(332).

The mesoscopic restructuring of the supporting surface does not come as a surprise for it is known from the studies of graphene growth on flat Ir(111) that terrace edges play an active role in graphene nucleation and growth and that, in fact, graphene islands reshape Ir steps during the growth [16]. Such substrate step-edge mobility seems to be general phenomenon for graphene growth on metal surfaces. For graphene which strongly binds to Ru(0001) surface, a creation of wider terraces and multisteps (facets) on substrate has been observed during graphene growth for the appropriate growth conditions [17]. Graphene covered wide (0001) terraces but did not grow over facets, which was explained by thermodynamically stable state of the surface consisting of large graphene covered areas and facets [17]. Note that compared to faceting on Ru(0001), graphene on Ir(332) exhibits formation of groups of several narrow steps, which are, in contrast to Ru facets, all covered by graphene. This difference might be due to the binding character of graphene to Ir and Ru substrates, i.e. it is weak van der Waals-like binding to Ir and strong chemisorption binding to Ru [18, 19].

Observed mesoscopic structure of graphene on Ir(332) could be generally described as an interplay of several parameters representing energy costs or benefits for the system. On the one hand, the major energy benefit in the restructured system comes from the van der Waals binding of graphene to created, large, flat (111) terraces. On the other hand, energy costs for the system include the energy needed to bend graphene over step bunches and to cause substrate restructuring, which is driven by temperature and ethylene decomposition chemistry.

Although our results indicate that graphene growth on Ir(332) leads to substrate restructuring, it does not necessarily mean that it is not possible to obtain graphene with very narrow TWD's over limited scales. Our results show that the growth temperature can be used to tune the TWD. We found that lower growth temperatures lead to comparably narrower TWD, while high growth temperatures of 930°C induce broad terraces with no obviously preferable terrace width (cf. Fig 1d and Fig. 3). Intuitively, this is because higher temperatures lead to higher mobility of surface atoms, and consequently, surface restructuring. For the lowest growth temperature of 770°C, TWD was centered around 2.03 nm and in the range between 0.5nm and 5.5nm. However, additional lowering of the growth temperature below 700°C is expected to lead to higher contribution of the dehydrogenated carboneous species [20], which is not suitable for our goal.

To get more insight on how lower synthesis temperatures produce graphene with more favorable TWD, we turned our attention to the analysis of moiré structures. LEED patterns obtained for graphene grown at 930°C (Fig. 4a), and 1120°C indicate significant contribution of the R30 graphene. The R30 contribution seems to be higher than is to be expected for similar growth preparation conditions on the Ir(111) surface [13, 16]. We believe this is due to much larger number of steps on Ir(332) which play a role in graphene nucleation, and which on flat (111) surfaces are not as nearly abundant [16]. Moreover, our STM data indicate that on Ir(332) the R30 and close rotations become much more abounded at lower temperatures which as a trend was also observed for graphene on Ir(111) [9].

From the analysis of STM images, we suggest that the structure of step edges, found at the surface, is on a microscopic scale related to rotational domains. Our STM images (cf. Fig. 4d) showed that the terraces, as well as the structural details at the terrace edges, such as kinks, are defined by the atop regions of the moiré structure of graphene. This is consistent with previous STM studies which showed that graphene covered Ir(111) terraces are always terminated by an integer number of atop regions [10], and that graphene islands and flakes as well consist of the integer number of atop regions which corner moiré cells. [21, 22]. Consequently, the moiré unit cell can be regarded as the smallest „building block" of graphene. The size of that building block is tuned by the rotation of graphene on iridium. The R30 rotation of graphene has substantially smaller moiré parameter than the R0 and the structural details, such as kinks, are therefore much smaller and less pronounced for the R30 graphene. Hence, one might expect that the size of the building block plays a role in an appearance of step edges, with seemingly straighter edges formed for smaller building blocks, for example for the R30 graphene.

It is indeed distinctive observation that the R30 graphene and close to it are accompanied by pronouncedly long and straight step edges, compared to the graphene orientations grouped around the R0 (cf. Fig. 4b). The average R30 domain has less step branching, and fewer, smaller kinks than the R0 domain. Correspondingly, these well defined terraces are a desirable feature in an attempt to impose an anisotropic periodic step potential which should affect the electronic band structure of graphene. All this suggest that using lower growth temperatures and forcing the growth of the R30 graphene on stepped Ir is a promising route to induce periodic modulation in graphene.

## 5. Conclusion

We have studied the growth on graphene on Ir(332) at different preparation temperatures using STM. Our results show that growing graphene at 770°C produces sample with relatively narrow TWD, where on average terraces are two times wider than that on an ideal Ir(332). Moreover, as a result of a rather low growth temperature, the areal contribution of the R30 graphene is substantial. As our analysis of the R30 areas indicates, graphene not only extends across Ir steps but also forms more uniform terrace edges. All this suggests that low growth temperatures (below 800°C) are advantageous for controlling the periodic structuring of graphene on Ir(332). Although single R30 domains, characterized in this work, are not extended over millimeter scales, which would be an optimal sample apply area-averaging techniques, areas of periodic features that we have formed should induce effects of a one dimensional periodic potential in the electronic band structure of graphene at well defined microscopic scales. This calls for further spectroscopic (scanning tunneling spectroscopy and μ-angle resolved photoemission spectroscopy) characterizations of this system in search for characteristic anisotropic effects in the band structure of graphene.

**Acknowledgements**

We thank M. Petrović and I. Delač Marion for useful discussion and help with the measurements, and T. Michely for valuable comments and suggestions. We gratefully acknowledge financial supports by the UKF (grant No. 66/10) and the MZOS (project No. 035-0352828-2840).

References:

[1] Geim K, Novoselov KS. The rise of graphene. Nat Mater 2007; 6:183-91.

[2] Avouris P, Chen Z, Perebeinos V. Carbon-based electronics. Nat Nanotec 2007; 2(1):605-15.

[3] Wintterlin J, Bocquet ML. Graphene on metal surfaces. Surf Sci 2009; 603:1841–52.

[4] Bae S, Kim H, Lee Y, Xu X, Park JS, Zheng Y, et al. Roll-to-roll production of 30-inch graphene films for transparent electrodes. Nat Nanotec 2010; 5:574-8.

[5] Park CH, Yang L, Son YW, Cohen ML, Louie SG. Anisotropic behaviours of massless Dirac fermions in graphene under periodic potentials. Nature Phys 2008; 4:213-7.

[6] Pletikosic I, Kralj M, Pervan P, Brako R, Coraux J, N'Diaye AT, et al. Dirac Cones and Minigaps for Graphene on Ir(111). Phys Rev Lett 2009; 102(5):056808-4.

[7] Rusponi S, Papagno M, Moras P, Vlaic S, Etzkorn M, Sheverdyaeva PM, et al. Highly Anisotropic Dirac Cones in Epitaxial Graphene Modulated by an Island Superlattice. Phys Rev Lett 2010; 105(24):246803-4

[8] van Gastel R, N'Diaye AT, Wall D, Coraux J, Busse C, Buckanie NM, et al. Selecting a single orientation for millimeter sized graphene sheets. Appl Phys Lett 2009; 95:121901-4.

[9] Hattab H, N'Diaye AT, Wall D, Jnawali G, Coraux J, Busse C, et al. Growth temperature dependent graphene alignment on Ir(111). Appl Phys Lett 2011; 98:141903-6.

[10] Coraux J, N'Diaye AT, Busse C, Michely T. Structural coherency of graphene on Ir(111). Nano Lett 2008; 8(2):565-70.

[11] Li A, Cai W, An J, Kim S, Nah J, Yang D, et al. Large-area synthesis of high-quality and uniform graphene films on copper foils. Science 2009; 324:1312-4.

[12] Preobrajenski B, Ng ML, Vinogradov AS, Mårtensson N. Controlling graphene corrugation on lattice-mismatched substrates. Phys Rev B 2008; 78(7):073401-4.

[13] Loginova E, Nie S, Thürmer K, Bartelt NC, McCarty KF. Defects of graphene on Ir(111): Rotational domains and ridges. Phys Rev B 2009; 80(8):085430-8.

[14] N'Diaye AT, Coraux J, Plasa TN, Busse C, Michely T. Structure of epitaxial graphene on Ir(111). New J Phys 2008; 10:043033-16.

[15] Horcas I, Fernandez R, Gomez-Rodriguez JM, Colchero J, Gomez-Herrero J, Baro AM. WSXM: A software for scanning probe microscopy and a tool for nanotechnology. Rev Sci Instrum 2007; 78:013705-8.

[16] Coraux J, N'Diaye AT, Engler M, Busse C, Wall D, Buckenie N, et al. Growth of graphene on Ir(111). New J Phys 2009; 11:023006-22.

[17] Günther S, Dänhardt S, Wang B, Bocquet ML, Schmitt S, Wintterlin J. Single terrace growth of graphene on a metal surface. Nano Lett 2011; 11:1895-1900.

[18] Busse C, Lazić P, Djemour R, Coraux J, Gerber T, Atodirese N, et al. Graphene on Ir(111): Physisorption with Chemical Modulation. Phys Rev Lett 2011; 107(3):036101-4.

[19] Wang B, Bocquet ML, Marchini S, Günther S, Wintterlin J. Chemical origin of a graphene moiré overlayer on Ru(0001). Phys Chem Chem Phys 2008; 10:3530-4.

[20] Lizzit S, Baraldi A. High-resolution fast X-ray photoelectron spectroscopy study of ethylene: From chemisorption to dissociation and graphene formation. Catal Today 2010; 154:68-74.

[21] Phark S, Borme J, Vanegas AL, Corbetta M, Sander D, Kirschner J. Atomic structure and spectroscopy of graphene edges on Ir(111). Phys Rev B 2012; 86(4):045442-4.

[22] Lu J, Yeo PSE, Gan CK, Wu P, Loh KP. Transforming C60 molecules into graphene quantum dots. Nat Nanotechnol 2011; 6:247-52.

**List of captions for Figures and Tables:**

**Table 1.** Overview of preparation parameters used in this work with a reference to figures with STM images.

**Figure 1.** a) Structural model of a (332) fcc-crystal surface generated by Surface Explorer software[1]. b) LEED pattern taken at the electron energy of 64 eV, c) STM image of clean Ir(332) with an inset showing 1.5 times magnified region marked by a white dashed rectangle, and d) TWD of clean Ir(332).

**Figure 2.** a) STM image of graphene area showing atomic and moiré resolution. b) STM image of a Ir(332) surface partially covered by graphene. Panels below are schematic models of graphene covered and bare Ir(332). TWD of c) graphene grown by CVD method at 775°C and d) graphene grown by TPG+CVD method at 770°C.

**Figure 3.** STM images of graphene covered Ir(332) for different preparation temperatures (a-c), and corresponding TWD's (d-f). Inset in c) shows 1.5 times magnified region marked by white dashed rectangle. The black line corresponds to the diagonal direction across the moiré unit cell.

**Figure 4.** a) LEED pattern taken at the electron energy of 59 eV, after growing graphene by TPG+CVD method at temperature of 930°C. b) STM image of two connected rotational domains of graphene corresponding to R0 and R30 groups. c) STM image of the R30 graphene showing atomic

resolution at terraces and step bunches where three kinks are marked by blue lines. d) STM image showing several domains within the R0 group graphene with clearly visible moiré patterns and four kinks marked by blue lines.